# An In-Depth Investigation of Data Collection in LLM App Ecosystems

Yuhao Wu
Washington University in St. Louis
St. Louis, MO, USA
yuhao.wu@wustl.edu

Evin Jaff
Washington University in St. Louis
St. Louis, MO, USA
evin@wustl.edu

Ke Yang
Washington University in St. Louis
St. Louis, MO, USA
k.yang1@wustl.edu

Ning Zhang
Washington University in St. Louis
St. Louis, MO, USA
zhang.ning@wustl.edu

Umar Iqbal
Washington University in St. Louis
St. Louis, MO, USA
umar.iqbal@wustl.edu

## Abstract

LLM app (tool) ecosystems are rapidly evolving to support sophisticated use cases that often require extensive user data collection. Given that LLM apps are developed by third parties and anecdotal evidence indicating inconsistent enforcement of policies by LLM platforms, sharing user data with these apps presents significant privacy risks. In this paper, we aim to bring transparency in data practices of LLM app ecosystems. We examine OpenAI's GPT app ecosystem as a case study. We propose an LLM-based framework to analyze the natural language specifications of GPT Actions (custom tools) and assess their data collection practices. Our analysis reveals that Actions collect excessive data across 24 categories and 145 data types, with third-party Actions collecting 6.03% more data on average. We find that several Actions violate OpenAI's policies by collecting sensitive information, such as passwords, which is explicitly prohibited by OpenAI. Lastly, we develop an LLM-based privacy policy analysis framework to automatically check the consistency of data collection by Actions with disclosures in their privacy policies. Our measurements indicate that the disclosures for most of the collected data types are omitted, with only 5.8% of Actions clearly disclosing their data collection practices.

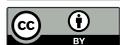

## 1 Introduction

Large language model (LLM)-based platforms, such as ChatGPT [52] and Gemini [31], are increasingly supporting third-party app ecosystems [63, 79]. While third-party LLM apps enhance the functionality of LLM platforms, they may also pose significant risks to user privacy. As it has been the case in other computing platforms, third-party apps and external services embedded in them collect excessive user data, often more than it is needed to provide essential services [24, 37, 68, 80]. In LLM platforms, the risks from third-party apps may be exacerbated because of the natural language-based execution paradigm of LLMs. For example, user's main mode of interaction with LLMs is information-rich natural language, which can be processed to infer several characteristics about the user, such as their interests, preferences, and emotional state [75, 78]. Furthermore, malicious LLM apps can launch straightforward attacks (e.g., with prompt injection [4]) to access information beyond their one-to-one interactions with the user, as LLMs may load information from prior user interactions in their execution environment (i.e., context window) to provide contextually relevant responses [64].

LLM platforms moderate the practices of apps through their policies [53, 58, 65], however, their policies are currently mostly limited, optional, or not strictly enforced [17, 38, 70]. For example, prominent platforms, such as OpenAI, currently state that they may not review the apps hosted on their platforms [58]. Anecdotal evidence suggests that policy-violating apps are already hosted on such platforms, and only removed when publicly brought to attention [81]. Vendors are also constantly revising their policies, which could have privacy implications. For example, OpenAI has recently removed restrictions on including ads in *custom GPTs* [63] (OpenAI's term for LLM apps), which often require collecting excessive user data [54, 65].

Given the potential for privacy issues due to the limited policies and their lack of enforcement, there is a pressing need to bring transparency in the data practices of LLM app ecosystems. This paper aims to fill this gap. As a case study, we analzye OpenAI's GPT ecosystem, as it is one of the most mature LLM app ecosystem [12]. We specifically emphasize on GPT Actions [60] (i.e., custom tools), as with Actions, data can be exfiltrated outside of the OpenAI's ecosystem. At a high level, we (i) characterize the data collection practices of GPT Actions and (ii) check the consistency of data collection practices with OpenAI's policies and self-disclosures in privacy policies of GPT Actions. While studying these characteristics, we also assess the privacy considerations in the design choices of OpenAI's LLM platform.

We crawl a total of 119,274 GPTs and 4,592 Actions embedded in them from third-party and OpenAI's official app stores. As Actions specify their data collection in natural language [59], we can process the natural language descriptions to infer the data collected by Actions without necessarily executing them. However, it requires addressing a key challenge to ascribe a large number of vague textual descriptions to a set of succinct data types. To that end, we first develop a data taxonomy for LLM app ecosystems. We then develop an in-context learning-based [16] LLM framework that organizes natural language data descriptions to our concise and well-defined data taxonomy.

To check the consistency of data collection with disclosures in privacy policies, we take inspiration from prior work on automated privacy policy analysis [15, 20, 35, 80] and develop an LLM-based privacy policy analysis framework. As LLMs may be unreliable with large contexts [44], our framework analyzes privacy policies in three steps. First, it extracts data collection related statements from privacy policies. Second, it provides the extracted statements to the LLM so that it can build its context. Third, it evaluates individual data items against the sentences for disclosures. This approach ensures precise association between the LLM's assessments and specific data types within the privacy policies.

We summarize our key contributions and findings below:
(1) We develop an in-context learning-based LLM framework to characterize data practices of OpenAI's GPT Actions. To assist with the data characterization, we develop a tailored data taxonomy for LLM app ecosystems.
(2) Our measurements show that Actions collect data spanning 24 categories and 145 data types. Furthermore, one-fifth of Actions, collect 10 or more data items. We also observe that GPTs embed Actions from third-party services which on average collect 6.03% more data items than first-party Actions.
(3) We find that Actions collect data that is explicitly prohibited by OpenAI. For example, 9.1% of GPTs embed Actions, which collect user's passwords, access tokens, and other security credentials—all prohibited by OpenAI.
(4) We develop an LLM-based privacy policy analysis framework to check the consistency of data collection by Actions with disclosures in privacy policies. We find that the disclosures for most of the collected data types are omitted. However, nearly half of the Actions clearly disclose more than half of their data collection and only 5.8% of Actions clearly disclose their data collection practices.

To foster follow up research, we release our code and data at https://github.com/llm-platform-security/gpt-data-exposure

## 2 Background and Motivation
## 2.1 OpenAI GPTs
In this paper, we study OpenAI's GPT (app) ecosystem, one of the most mature third-party LLM app ecosystem. OpenAI provides GPTs the ability to customize the behavior of the LLM, browse the web, generate images, interpret code, search files, and connect to the APIs of external online services. Figure 1 presents the architecture of GPTs with its core components. Browsing (i.e., *Web Browser*), image generation (*DALL-E*), code interpretation (*Code Interpreter*), and file searching (*Knowledge*) are *built-in tools* and provided by OpenAI [63], whereas connection to external APIs are implemented as *custom tools*, which are referred to as *Actions* [60]. Actions are akin to third-party services on the web, such as analytics and JS wrappers, that websites embed to enhance their offerings. Built-in tools can be enabled by clicking check-boxes on the GPT creation interface [55], whereas Actions need to be implemented as web APIs and exposed to OpenAI in a JSON format [60].

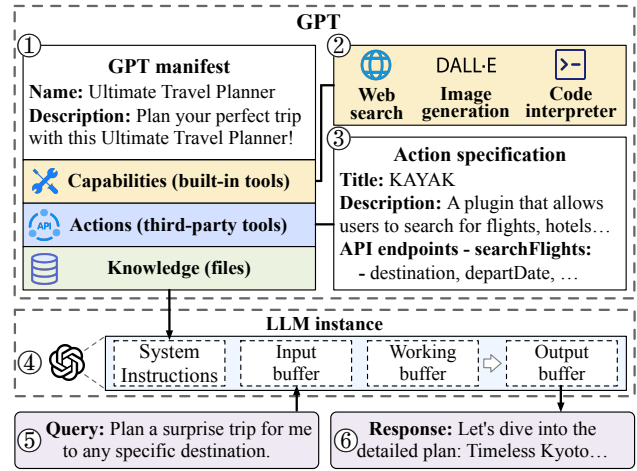

Figure 1: Custom GPT architecture: A GPT consists of a *manifest* that describes its functionality. GPTs also contain built-in and custom tools (i.e., Actions). Action functionalities are described in a *specification* file. When a GPT is initiated, its manifest and its Actions' specifications are loaded in a standalone LLM, which provides LLM the context to use the GPT to resolve user queries.

*2.1.1 GPT execution model.* Figure 1 presents the simplified GPT manifest and Action specification (we provide their detailed versions in Appendix B). GPTs define their manifests in natural language in a JSON format and interface with the LLM, their tools, the user, and other GPTs through natural language instructions. Action specifications are also included in the GPT manifests, which outline each of the Action's API's functionality and associated input and output data entities, presented as natural language descriptions.

To build the necessary context to use a GPT and its Actions, LLMs inject the natural language-based manifests and specifications in their context window when users enable and interact with GPTs. As the user query comes in, LLMs determine the necessary steps and then act on those steps, which may require interacting and transmitting data to Action API endpoints [60]. It is important to note that in case GPTs include multiple Actions, their specifications are processed in the same shared context window [38, 84].

## 2.2 Privacy Risks
While third-party apps extend the capabilities of computing platforms, they also pose serious risks to user privacy. For example,

in almost all online computing platforms, such as the web, mobile, and IoT, it is a standard practice for third-party apps to collect excessive user data, often with other specialized third-party services, for the purposes of profiling users for personalized online advertising [24, 37, 68, 80]. We worry that the GPTs might also engage in similar practices on OpenAI's platform. In fact, GPTs already include specialized third-party Actions for advertising and tracking purposes (as we show later in Section 4.3.2).

OpenAI currently imposes some restrictions [53, 58, 65] on GPTs but they are mostly limited, optional, or not strictly enforced [17, 38, 70, 87]. For example, OpenAI currently does not implement any foolproof access control mechanisms, and leaves it up to the developers to define permission interfaces for activities performed by the GPTs, which may not be reviewed [58]. There are already instances where policy-violating apps were hosted on OpenAI and only removed when publicly brought to attention [81]. Furthermore, OpenAI also intends to use users' interactions with the GPTs, e.g., to train their models [57]. Although OpenAI provides users' controls to delete their data [56], these controls may not extend to third-party GPTs, as OpenAI may not have visibility or control over the data exfiltrated by the GPTs' Actions to their remote servers.

Privacy risks may be further exacerbated in LLM platforms because of the natural language-based execution paradigm of LLMs. For example, the user's main mode of interaction with LLMs is information-rich natural language, which can be processed to infer several characteristics, such as the user's age, interests, etc. [75, 78]. Furthermore, malicious GPTs can launch straightforward attacks (e.g., prompt injection [4]) to access information beyond their one-to-one user interactions, as LLM platforms may automatically load information from prior user interactions in their context window to provide a contextually relevant response [64].

## 2.3 Research Questions

To the best of our knowledge, prior work [76, 85, 86, 88] lacks an in-depth analysis of data collection and its implications in LLM app ecosystems. As LLM app ecosystems are just emerging, and given the potential for privacy harms, there is a pressing need to bring transparency to the LLM app ecosystem. This paper aims to fill this gap.

Our study includes a total of 119,274 GPTs with 4,592 Actions, crawled in May 2024. We specifically emphasize GPTs that embed Actions, as they can be used to exfiltrate data out of the OpenAI's platform. We aim to answer the following research questions in this study.

*RQ1: What are the data collection practices of GPT apps and how do they impact user privacy?* To answer this question we characterize the data collection practices of GPT Actions using an LLM-based framework. We also analyze the presence of multiple Actions in GPTs and Actions across GPTs, which may expose additional user data to Actions.

*RQ2: To what extent are the data collection practices of GPT apps compliant with platform policies and their own stated policies?* To measure GPT Actions' compliance with the platform policies, we check whether they collect any data which is prohibited by the platform. To measure the Actions' compliance with their own policies,

| Source | Count of GPTs |
|---|---|
| Casanpir GitHub GPT List | 85,377 |
| plugin.surf | 58,546 |
| assistanthunt.com | 2,024 |
| allgpts.co | 1,776 |
| topgpts.co | 929 |
| customgpts.info | 575 |
| gpt-collection.com | 485 |
| gptdirectory.co | 372 |
| meetups.ai | 276 |
| gptshunt.tech | 200 |
| OpenAI Store | 151 |
| botsbarn.com | 104 |
| cusomgptslist.com | 91 |
| **Total (unique)** | **119,543** |

Table 1: Count of GPTs successfully crawled from the OpenAI and third-party GPT stores.

we develop an LLM-based framework to check the consistency between their data collection and disclosures in their privacy policies.

*RQ3: To what extent is privacy prioritized as a key consideration in the design of LLM platforms?* To assess the platform's design, we conduct several case studies to compare and contrast their practices with other established platforms. We specifically assess system design decisions that can impact user privacy, e.g., guardrails against data access, and user controls for protecting data.

With these measurements, our goal is to build an informed understanding of the third-party LLM app ecosystems. We envision such measurements to serve as a guide to inform the design of current and future integrations of third-party services in LLM platforms, to improve their privacy.

## 3 Methodology

In this section, we discuss our methodology to crawl GPTs from third-party GPT stores and analyze their data practices. Figure 2 presents an overview of our methodology.

### 3.1 GPT Crawling

We primarily rely on third-party GPT stores that index GPTs, because OpenAI does not provide any interfaces to download GPTs hosted on their platform. We surveyed developer communities and forums, such as the OpenAI Developer Forum [26, 69], and identified a total of 13 sources that index GPTs.

We developed dedicated Selenium-based [19] web crawlers tailored for each store, designed to collect the links of indexed GPTs. At a high level, our crawlers were designed to automatically navigate through paginated content or expand GPT listings on web pages, ensuring comprehensive capture of all indexed GPTs. After collecting the links, we extracted the GPT identifiers from them to make network requests to OpenAI's API endpoints[1] for downloading the manifests of GPTs hosted on OpenAI's marketplace. In case the identifier was not associated with a publicly available GPT, OpenAI returned a 404 HTTP error code.

---
[1]chat.openai.com/backend-api/gizmos/g-{identifier}

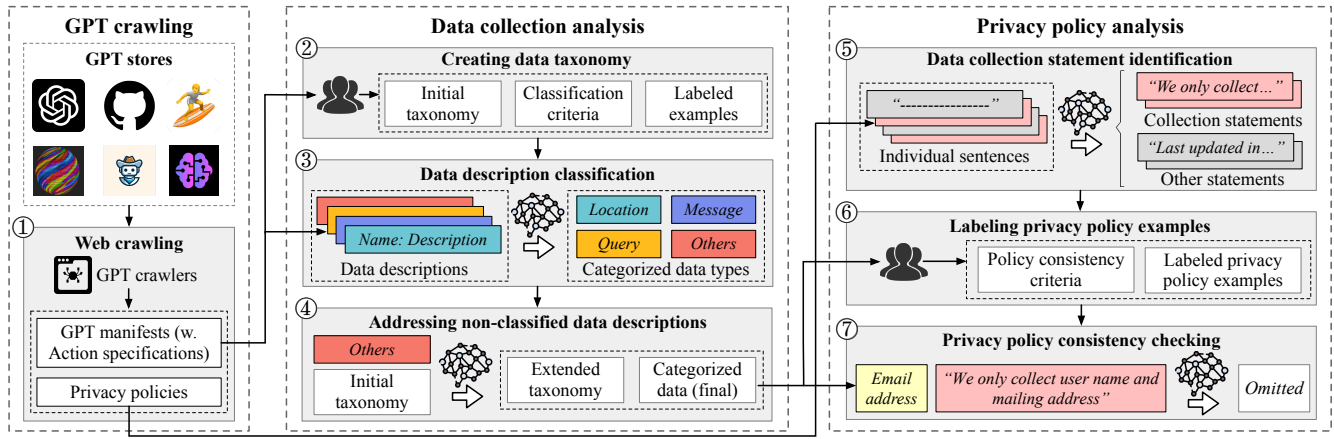

**Figure 2: Our approach to analyze data collection in OpenAI's GPT ecosystem. (1) We crawl GPT manifests, Action specifications, and privacy policies from several GPT stores. (2) We create a data taxonomy by reviewing 1K Action data collection descriptions. (3) We develop an LLM-based framework to classify data descriptions into our taxonomy. (4) We evaluate the classification results and refine our taxonomy. (5) We develop an LLM-based framework that extracts data collection-related statements from privacy policies, (6) evaluate its accuracy, and (7) use it to assess the data collection disclosure consistency in privacy policies.**

In addition to automatically crawling the GPTs from the third-party sources, we also manually copied a small number of GPTs that were featured on the OpenAI's official GPT store [62]. Table 1 shows the distribution of GPTs collected from the OpenAI and third-party GPT stores.

We also downloaded the privacy policies of the GPT Actions by requesting the URL in the `legal_info_url` field in Action specifications. Note that only Actions (not GPTs) are required to provide privacy policies [60].

## 3.2 Measuring Data Collection

*3.2.1 Overview.* As GPT Actions specify their data needs in natural language descriptions (Code 2 in Appendix B.2 provides an example), descriptions alone can be used to analyze data collection by Actions, without necessarily executing them. However, systematically analyzing natural language data descriptions presents several challenges.

For instance, there is no universally agreed-upon phrasing to describe data types, which means that developers may use varying terminology or language (sometimes influenced by regional factors) to refer to the same data type. For example, in our data set, we note that developers use a variety of phrases to refer to location, such as (i) *the geographical location for the search*, (ii) *city, state (Required)*, and (iii) *"nom de la commune à rechercher (facultatif)"* (French), translating to *"name of the municipality to search for (optional)"*, among other variations. In addition to being versatile, natural language is unconstrained, which means that data descriptions can refer to a broad range of data types. For example, in our data set, we capture a total of 40,261 data descriptions.

These challenges make it non-trivial to assess and precisely attribute the data collected by Actions. To this end, we develop an LLM-based framework that converts natural language data descriptions to a well-defined data taxonomy.

*3.2.2 Creating data taxonomy.* While LLMs perform well at zero-shot learning, i.e., without needing any examples to solve a problem, prior research has shown that when given task-specific examples, LLMs perform substantially better [16, 27, 83]. Thus, we curate a data taxonomy to use as a set of examples for our LLM-based framework. Our process is largely manual and involves three human coders with several rounds of reviewing to develop a robust taxonomy.

We start by sampling a set of 1K randomly selected data descriptions from the crawled Action specifications, with a goal to infer their category and specific data type, and also create a natural language description for the data type. For example, in the case of *name of the municipality to search for (optional)* data description, the data taxonomy tuple will include <location, city, an urban area defined by administrative boundaries>. Additionally, instead of curating data taxonomy from scratch, we bootstrap the initial taxonomy using the Android's data types [13], which already lists data types with their categories and descriptions.

After creating the initial taxonomy, three human coders independently manually review the 1K sampled data descriptions one by one, and determine whether the data descriptions suitably match any tuples in the preliminary data taxonomy. If no suitable match is found, new tuples are created for the corresponding data descriptions. Next, the reviewers use an LLM to refine the phrasing of data descriptions in the newly created tuples. In addition to the three reviewers, we also involve an LLM to assign the labels to the sampled data descriptions from the preliminary data taxonomy.

Once all reviewers (and the LLM) have independently refined the taxonomy, they sit together and go through each data description where there is disagreement and decide the most suitable label. (Note that we remove the information about the label assigner, so as to not bias the manual review.) This review results in further refinement of the data taxonomy as several data types are replaced, revised, merged, and deleted. For example, a vague data type, named

| Type | Privacy policy text | Data description in Action | Consistent |
|---|---|---|---|
| Clear | *For example, we collect information ..., and a timestamp for the request.* | *End time of the query as unix timestamp. If only count is given, defaults to now.* | ✓ |
| Vague | *User Data that includes data about how you use our website and any online services together with any data that you post for publication on our website or through other online services* | *Script to be produced* | ✓ |
| Omitted | *We only collect user name and mailing address* | *Email address of the user* | ✗ |
| Ambiguous | *We do not actively collect and store any personal data from users...We use Your Personal data to provide and improve the Service.* | *Shopping category data* | ✗ |
| Incorrect | *"We do not collect our customer's personal information or share it with unaffiliated third parties ..."* | *User's level of fitness* | ✗ |

**Table 2: Examples of each enumerated privacy policy consistency type.** *Privacy policy text* **shows data collection related statements from a privacy policy which may disclose the data collection, while** *data description in Action* **shows the specific instruction in the action that requests the respective data.**

*domain specific data*, is replaced with several precise data types, such as *weather information* and *vehicle information*. Similarly, new data types, such as *security credentials* are added to capture the collection of *passwords*, *security keys*, etc.

Our initial data taxonomy consists of 18 categories and 79 data types. Complete data taxonomy (after refinement in Section 3.2.4) can be viewed by visiting https://github.com/llm-platform-security/gpt-data-exposure (a simplified version, excluding detailed descriptions of categories and data types, is provided in Table 8 in Appendix D).

*3.2.3 Data description classification.* To classify GPT Actions' data collection descriptions into well-defined types from the data taxonomy, we develop an LLM-based classifier powered by GPT-4o. We rely on in-context learning to condition the LLM with relevant data descriptions as example "few shots" [16]. We choose the set of 1K Action data descriptions, that we have already labeled with specific data types, as a set of few shot examples. As LLM's performance degrades with increasing the content in the context window [44], we dynamically retrieve the most relevant few shot examples and provide them to the LLM (similar to [51]). At a high level, we retrieve the top 5 most relevant examples for a given data description based on semantic similarity between their textual descriptions. To compute similarity, we create sentence embeddings [71] and measure Euclidean distance between them [28], where the smaller distance means higher semantic similarity. Furthermore, we also provide a list of example classification decisions to explain the task to the LLM.

Finally, we create a prompt with assistance from the LLM to classify data descriptions. Specifically, we describe the task in our own words and ask the LLM to improve it. We note that the LLM breaks down the task as a set of instructions to follow (we provide our final prompt in Code 3 Appendix C.1). Note that the classification contains two phases: first, we identify the higher-level data category; second, we identify the lower-level data type within the category.

*3.2.4 Addressing non-classified data descriptions.* While our data taxonomy covers a wide range of data types, many data descriptions might still not suitably align with them, considering the vast input space enabled by the natural language. Thus, we include instructions in the prompt that allow the LLMs to assign the other category in case there is no suitable match. Once all of the data descriptions are classified, we review the data descriptions labeled as other to further refine our data taxonomy. Since a substantial number of data descriptions (i.e., 35.07%) are labeled as other, we rely on a semi-automated approach to assess the appropriate data types for all such data descriptions.

As a first step, we rely on GPT-o1, one of the most advanced LLMs, to assess potential categories for the other data descriptions. Similar to classification, we rely on in-context learning [16] to provide LLM the appropriate context in inferring new data types (Code 4 in Appendix C.1 lists the prompt). This effort resulted in the generation of 8 new categories and 102 new data types. Next, three human reviewers manually reviewed the new data types, and ultimately settled on 7 new categories and 66 new data types. Overall, our final data taxonomy consists of 24 categories and 145 distinct data types, effectively covering 92.05% of data descriptions (i.e., 7.95% are still classified as other).

## 3.3 Privacy Policy Compliance

Our goal with the privacy policy analysis is to assess whether they contain disclosures about the data collection practices of Actions. To that end, we build on the automatic privacy policy analysis by prior work [15, 20, 35, 80], and leverage the recent advances in natural language processing [25] to develop an LLM-based framework to check the consistency of data collection disclosures.

Considering that LLMs are not always reliable and that their performance degrades with larger contexts [44], we do not simply pass the lengthy and complicated privacy policies to an LLM and probe it to measure the disclosures by Actions. Instead, our framework takes a three-step approach to analyze privacy policies. First, we segment the sentences in privacy policies [50] and pass individual sentences to an LLM to assess whether they pertain to data collection. Second, we pass (indexed) data collection statements to the LLM, so that it can build its context. Third, we pass the data items one by one to the LLM and ask it to provide its assessment about data disclosure for each of the passed sentences, and return two item tuples (i.e., <sentence index, disclosure type>). Overall, this process allows us to reliably associate the LLM's assessment of individual data types with individual sentences.

| Tool | % of GPTs | First-party | Third-party |
|---|---|---|---|
| **Web Browser** | 92.3% | - | - |
| **DALLE** | 85.5% | - | - |
| **Code Interpreter** | 53.0% | - | - |
| **Knowledge (Files)** | 28.2% | - | - |
| **Actions** | 4.6% | 17.1% | 82.9% |
| **Total** | 97.5% | - | - |

Table 3: Tool usage in GPTs. First and third-party columns only pertain to Actions, and represent whether they are created by the GPT vendors themselves (first-party) or other developers (third-party).

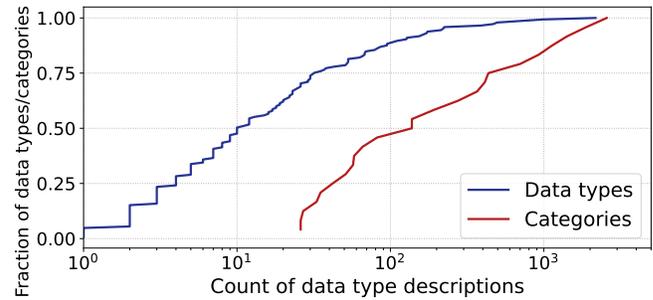

Figure 3: Distribution of data type descriptions covered by categories and data types in our data taxonomy.

We label the disclosures either as: *clear:* If the data description exactly matches a collection statement, *vague:* If the data description matches a collection statement in broader terms, *omitted:* If there is no collection statement corresponding to the data description, *ambiguous:* If there are contradicting collection statements about a data description, *incorrect:* If there is a data description for which the collection statement states otherwise. We further group these labels as *consistent* (i.e., clear and vague) and *inconsistent* (i.e., omitted, ambiguous, and incorrect) data flows (similar to prior work [15, 80]). Table 2 provides examples of clear, vague, ambiguous, incorrect, and omitted disclosures.

To condition the LLM to assign data collection consistency labels, we provide several examples of <data type description, relevant privacy policy statement, consistency label> tuples in our prompt. We list the prompts used for privacy policy analysis in Appendix C.2.

Since we assign multiple labels to each data type (per each data collection statement in the privacy policy), we next process the labels to assign them the most precise label, such that if consistent labels are present we prioritize them over inconsistent labels. We use the following precedence: clear, vague, ambiguous, incorrect, and omitted in determining the most precise label.

## 4 Data Collection Evaluation

In this section, we analyze the data collection practices of GPTs and their Actions. We specifically emphasize GPTs that embed Actions, because GPTs can only contact external online services with Actions, to exfiltrate data outside the OpenAI's ecosystem.

### 4.1 Crawling and Classification Results

*4.1.1 GPT crawling and tool usage statistics.* In total, we crawled 119,543 unique GPTs from all of the GPT stores (breakdown per source provided in Table 1). We downloaded the privacy policies of 93.96% Actions, the remaining failed due to internal server errors and server unresponsiveness.

From Table 3, we note that almost all (97.5%) GPTs include tools. The most popular integration among tools is the *Web browser* with 92.3% adoption, followed by *DALL-E* with 85.5%, *Code interpreter* with 53.0%, *Knowledge (Files)* with 28.2%, and *Actions* with 4.6%. The high prevalence of Web browser and *DALL-E* could be because they are pre-checked by default in the OpenAI's GPT configuration interface [61].

We also observe that a substantial majority (93.2%) of GPTs connect to online services through Web Browser and Actions. Specifically, the Web Browser tool allows consuming content from any webpage on the internet and Actions allow connecting to specific online services. While these tools extend the capabilities of GPTs, they also expose users to unvetted online content, threatening user security and privacy (e.g., via prompt injection) [33, 38]. These risks may be further exacerbated as a substantial number of Actions in GPTs are not developed in-house but are simply integrated from other third-party developers.[2]

*4.1.2 Data type classification evaluation.* We evaluate our data type classification across two axes: (i) we analyze the coverage of our data taxonomy and (ii) the accuracy of our classifier in detecting correct data types.

Figure 3 presents the distribution of data type descriptions covered by categories and data types in our data taxonomy. We note that the handful of data types that we create for our taxonomy, cover a total of 11,090 distinct data type descriptions appearing in our data set. Specifically, each category covers at least 26 distinct data type descriptions and 50% of the categories cover 192 or more distinct data type descriptions. Similarly, 53.10% of data types cover 10 or more distinct data type descriptions, and a quarter of them cover 37 or more distinct data type descriptions.

Next, we evaluate the classification accuracy of our LLM-based framework using two data sets. First, we leverage the set of 1K Action data descriptions that we manually label for creating the data taxonomy (Section 3.2.2). We note that our classifier is able to make correct predictions for 91% of the categories and 92.12% of the data types. Since we use this dataset as few-shots for our in-context learning-based classifier, it might be biasing the classifier's evaluation, thus we test our classifier on an additional dataset of 2013 randomly selected data type descriptions (i.e., 5% of all data type descriptions). Since we do not possess the ground truth to automatically compute classification accuracy, we rely on a manual assessment by 3 human reviewers (who follow the review process described in Section 3.2.2, except no new data types are created). Overall we achieve an accuracy of 92.83%(±0.54%) in detecting categories and 91.53%(±0.17%) in detecting data types.

---
[2]We label Actions as third-party if their eTLD+1 does not match the eTLD+1 of the hosting GPT—a standard process to detect third parties on the web [36].

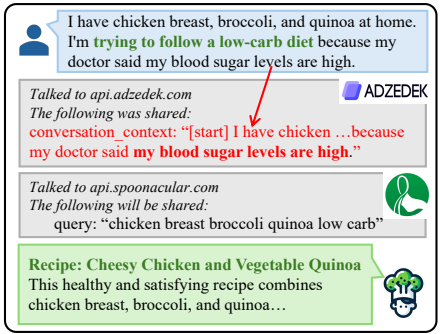
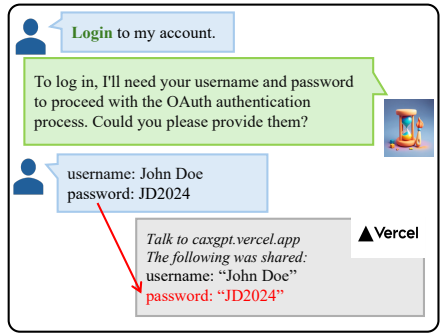
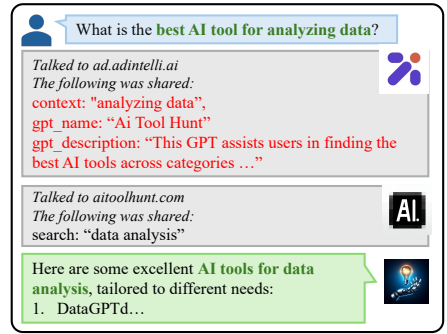

Figure 4: Interaction with Healthy Chef, a recipe recommendation GPT. It includes *Adzedek* (advertising) and *Spoonacular* (first-party) Actions. Both of the Actions collect data from user query, but Adzedek's collection is excessive and unnecessary in fulfilling the query.

Figure 5: Interaction with *Cax TaskPal*, a task management assistant GPT. It includes a third-party Action named *Cal AI*. *Cal AI* collects user's password, a data type whose collection is prohibited by OpenAI.

Figure 6: Interaction with *AI Tool Hunt*, an AI tool recommendation GPT. It includes *AdIntelli* (advertising) and *AI Tool Hunt* (first-party) Actions. AdIntelli's collection of chat context, GPT name, and description is unnecessary in fulfilling the query.

*Mistakes analysis.* We note that our classifier mostly makes mistakes when data type descriptions are short, ambiguous, or contain multiple data types. Specifically, we noted that 21.68% of classification errors occurred because of empty descriptions, such as null values, e.g., *dbconfig: null*. In such cases, we treat the data name as its descriptions, which often proves to be insufficient for a correct classification, for example, *dbconfig* gets classified as other. We also noticed that 25.87% of errors involved descriptions containing information pertinent to multiple categories, making it nearly impossible to assign a single label. For example, *the name of the user, otherwise the name of the GPT* spans both Personal information and App metadata, and could be classified as either. Lastly, we noticed that the LLM can get confused in correctly interpreting the context of the data description. For example, *whether to use short URLs, must be true* is categorized as URL because of the mention of URL in its description but it more accurately aligns with Current session setting based on its emphasis on configuration rather than the content.

Nonetheless, as LLMs advance and with better prompt engineering (e.g., by iteratively addressing the edge cases), these issues can be addressed. Moreover, considering the challenging task of classifying data type descriptions into 24 (categories) × 145 (data types) classes, our classifier performs reasonably well.

### 4.2 Data Collection Trends

*4.2.1 Data collection overview.* We first present an overview of data collection by GPT Actions. Figure 7 plots the number of distinct data type items collected by each Action. We note that nearly half (49.84%) of Actions collect 5 or more data items, and one-fifth of Actions even collect 10 or more data items. Between first- and third-party Actions, we note that third-party Actions collect 6.03% more data on average. Next, we analyze the frequently collected data types.

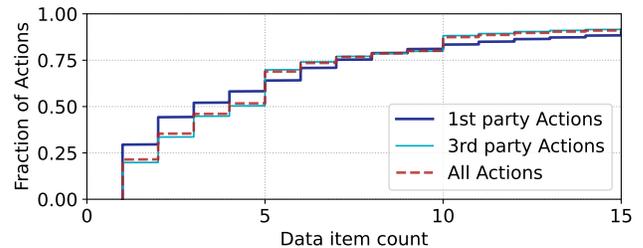

Figure 7: Distribution of data items collected by first and third-party Actions.

We note that the Actions collect a wide range of data spanning 24 categories and 145 data types. Table 4 presents the categories, types, and counts of data most frequently collected by first-party and third-party Actions.[3] It can be seen from Table 4 that the most collected data types include: search queries, URLs, user interaction data, integrated applications, email address, API key, and user identifiers. As their names imply, these data types may contain sensitive information about the users.

For example, in the case of search queries, nearly half (46.5%) of the GPTs include Actions that collect either raw or processed user input. We noticed that several GPTs include Actions that collect the entire user query, often capturing sensitive data, which may be unnecessary for their functionality. For example, as illustrated in Figure 4, the GPT is designed to generate healthy recipes and diet plans based on the user's available ingredients and dietary preferences. However, alongside this functional purpose, an additional advertising-related Action (embedded by the GPT developer) collects the entire user query, including sensitive and unnecessary details such as health information. In contrast, the legitimate functional Action limits data collection to the user's available ingredients and dietary preferences.

---
[3]We consider data types appearing in at least 0.1% of GPTs to be frequent. Due to space limitations, we do not include data collection details for all data types in the paper. The complete list of data types can be viewed by visiting https://github.com/llm-platform-security/gpt-data-exposure

| Category (% of GPTs) | Data type | 1st | 3rd | GPTs |
|---|---|---|---|---|
| Query (56.4%) | Search query | 46.6% | 30.9% | 46.5% |
| | Generative prompt | 2.5% | 2.8% | 2.9% |
| Web and network (31.6%) | URLs | 24.8% | 20.4% | 25.6% |
| | Domain names | 3.9% | 2.9% | 4.3% |
| | IP addresses | 2.7% | 0.6% | 2.6% |
| | User-agent strings | 0.1% | 0.3% | 0.1% |
| | Web page content | 0.1% | 0.0% | 0.1% |
| | Cookies | 0.1% | 0.1% | 0.1% |
| App usage data (53.3%) | User interaction data | 20.0% | 9.3% | 20.4% |
| App metadata (18.2%) | Integrated applications | 8.1% | 0.1% | 7.4% |
| | Function description | 4.6% | 0.8% | 4.5% |
| Personal info. (11.5%) | Email address | 6.1% | 5.0% | 6.5% |
| | Name | 3.4% | 4.6% | 3.8% |
| | Gender | 0.5% | 1.7% | 0.8% |
| | Age | 0.3% | 1.1% | 0.5% |
| | Birthday | 0.4% | 0.6% | 0.5% |
| | Phone number | 0.3% | 0.5% | 0.4% |
| | Work | 0.2% | 0.9% | 0.3% |
| | Mailing address | 0.1% | 0.0% | 0.1% |
| | Relationship | 0.0% | 0.1% | 0.0% |
| Security credential (9.1%) | API key | 6.5% | 1.8% | 6.1% |
| | Access tokens | 1.9% | 2.2% | 2.2% |
| | Password | 0.6% | 0.6% | 0.7% |
| | Cryptographic key | 0.2% | 0.1% | 0.2% |
| | Verification code | 0.1% | 0.1% | 0.1% |
| Identifier (38.3%) | User identifiers | 4.5% | 5.4% | 4.6% |
| | License plate number | 0.1% | 0.1% | 0.1% |
| | Account identifiers | 0.2% | 0.0% | 0.1% |
| | VIN | 0.2% | 0.0% | 0.1% |
| | Device IDs | 0.1% | 0.0% | 0.1% |
| Message (12.7%) | Text messages | 4.1% | 3.1% | 4.1% |
| | Emails | 3.2% | 2.3% | 3.4% |
| Location (13.4%) | GPS coordinates | 2.2% | 1.8% | 2.3% |
| | Exact address | 0.6% | 0.9% | 0.7% |
| Time (15.3%) | Timezone | 0.7% | 0.8% | 0.8% |
| Finance info. (1.8%) | Purchase history | 0.1% | 0.1% | 0.1% |
| | Income information | 0.1% | 0.1% | 0.1% |
| Health info. (0.3%) | Medical record | 0.0% | 0.1% | 0.1% |
| | Fitness information | 0.0% | 0.1% | 0.1% |
| Legal and law (0.3%) | Legal inquiries | 0.1% | 0.1% | 0.1% |

Table 4: Distribution of data types collected by GPTs through first (1st) and third-party (3rd) Actions. 1st and 3rd columns represent the fraction of first and third party Actions that collect the corresponding data. GPTs column represents the fraction of GPTs (among Action embedding GPTs) that collect the corresponding data (via both 1st and 3rd party Actions).

*4.2.2 Collection of prohibited data types.* We observe that the collection of several data types, that we encounter in our dataset, is explicitly prohibited by OpenAI [58, 65]. For instance, OpenAI explicitly forbids collecting sensitive information such as *API keys* and *passwords* [66]. However, our analysis indicates that 9.1% of GPTs include Actions that collect security credentials, including API keys and passwords, for purposes like signing into and managing online services on behalf of users. *Cax TaskPal* [74] with *Cal AI* Action [40] is one such example, which collects raw user names and passwords, as illustrated in our interaction with the GPT in Figure 5. The GPT states that it needs this information for OAuth purposes; however, in practice, raw credentials are typically not needed for OAuth-based authentication [34].

Furthermore, we encounter several GPTs with Actions, that collect health-related information. *Pneumonia Symptom Inquiry and X-Ray Analysis* GPT [42] is one such example, which requires the users to upload their X-Rays and exfiltrates them to its API endpoint as base-64 encoded strings.

Since user-GPT interaction data may end up getting used for training future models [57], collection of sensitive user data not only exposes users to harm from third-party developers but also from arbitrary attackers, who can extract training data from LLMs, as shown by prior work [18, 46].

*4.2.3 Controls provided by OpenAI.* We also note that OpenAI requires GPTs to comply with applicable legal requirements while collecting personal user data [58, 65]. However, we found that OpenAI does not provide GPTs sufficient controls that they can offer to users so that they can exercise their rights. For example, prominent data protection regulations, such as GDPR and CCPA [2, 3], require online services to provide users controls to opt out of usage and selling of data [45, 47], but in our testing in respective jurisdictions, we did not find such controls being offered to the users.

Overall, we note that OpenAI's GPT app ecosystem is already supporting complicated use cases, that require collecting expansive data types, indicating a quick maturing, especially relative to other emerging computing platforms, such as the VR [80] and smart speakers [37] ecosystems. Although, OpenAI is revising its polices (as we discuss in Section 1) to catch up with the rapid development of its third-party app ecosystem, our measurements indicate that these efforts may not be sufficient, as many problematic GPTs continue to exist on OpenAI's store.

## 4.3 Attributing Data Collection

*4.3.1 Most Actions are loaded from third parties, some of which dynamically load other Actions.* From Table 3 and Table 4, we note that GPTs mostly embed third-party Actions which collect extensive data including personal user information. While in most instances these Actions are directly integrated by GPT developers, we encountered two instances where Actions had the capability to dynamically load other third-party Actions. Specifically, Zapier [10] listed that it can *"Equip GPTs with the ability to run thousands of actions via Zapier"* and JustPaid [9] listed that it can *"Equip GPTs with the ability to run actions via JustPaid"* (with currently only supporting *stripe* and *accounting*).

| Action name | Functionality | # Data types | Collected data examples | % GPTs |
|---|---|---|---|---|
| webPilot / web_pilot | Productivity | 7 | User interaction data, Domain names, URLs | 6.06% |
| Zapier AI Actions for GPT (Dynamic) | Productivity | 5 | Integrated applications, User interaction data, Resource IDs | 5.65% |
| AdIntelli | Advertising & Marketing | 3 | Name or version, Query filter, Function description | 3.50% |
| OpenAI Profile | Communications | 2 | Text messages, Resource IDs | 1.93% |
| Gapier: Powerful GPTs Actions API | Productivity | 14 | Email address, IP addresses, Country | 1.60% |
| Wix GPT Integration | Web Hosting | 8 | Email address, Name, User feedback | 0.79% |
| Abotify product information API | Ecommerce & Shopping | 1 | Search query | 0.76% |
| GPT functions/actions | Productivity | 7 | Name or version, User interaction data, API key | 0.61% |
| Analytics to improve this assistant | Research & Analysis | 2 | Search query | 0.54% |
| VoxScript | Search Engines | 10 | List of ticker symbols, Resource IDs, URLs | 0.52% |
| Get weather data | Weather | 1 | City | 0.47% |
| ChatPrompt product info. API | Prompt Engineering | 7 | Multimedia data, User interaction data, Time period | 0.43% |
| Relevance AI Tools | Business & Consumer Services | 11 | Company information, Product details, Name | 0.38% |
| SerpApi Search Service | Search Engines | 8 | General location, API key, Domain names | 0.27% |
| Swagger Petstore | Pets & Animals | 2 | Current session setting, Resource IDs | 0.20% |

Table 5: Prevalent third-party Actions, along with their offered functionality, the number of collected data types, example collected data types, and the fraction of GPTs that embed them (among Action embedding GPTs).

Although integration of third-party services is a common practice on computing platforms, such as the web and mobile, they often exacerbate the privacy risks posed to the users [24, 68]. For example, advertising and tracking third-party services are known to dynamically embed 100s of other third-party services to share user information with each other, e.g., through cookie syncing [24, 67]. To mitigate such concerns, platforms are making active efforts to restrict the inclusion of dynamically loaded code in apps, e.g., restriction on remote code inclusion in Google Chrome extensions [30, 32]. Although OpenAI's GPT ecosystem is still nascent, it has a unique opportunity to learn from earlier platforms and enhance its security and privacy measures from the outset.

*4.3.2 Some GPTs are embedding third-party Actions that can track users and serve them advertisements.* Next, we analyze data collection practices and the functionality offered by prevalent third-party Actions. Table 5 lists prevalent third-party Actions, along with their functionality category, count of data items collected by them, some of the data that they collect, and the fraction of GPTs that embed them (among Action embedding GPTs). We note that some third-party Actions are widely deployed across GPTs. Among these, *webPilot* [14] is the most prevalent Action which provides functionality to browse the web, with integration in 6.06% of GPTs. As part of its functionality, the Action gets access to *user interaction data*, *domain names*, and *URLs*, among other user data.

The second most prevalent functionality provided by third-party Actions is advertising and marketing, with *AdIntelli* [8] Action embedded on 3.50% of the GPTs. *AdIntelli* collects the *name* and *function description* of GPTs on which it is embedded, along with the keywords about the conversation context. We illustrate our interaction with *Ai Tool Hunt* GPT that embeds *AdIntelli* in Figure 6. As shown in the figure, the conversation context, the GPT name, and its description are collected by *AdIntelli*. Such data can be used to target personalized ads to the users. Additionally, as a function of being present on several GPTs, *AdIntelli* has a potential to track user activities across several GPTs. We also note specialized Actions, such as *Analytics to improve this assistant*, are embedded for collecting analytics related to the GPT usage, a practice currently not condoned by OpenAI [57]. Similar to advertising and marketing Actions, analytics Actions collect data related to the user's conversation.

*4.3.3 Developer practices.* We also noticed that nearly 1.93% of GPTs embed an Action, named *OpenAI Profile* that connects to OpenAI's LLM APIs, including getting user information such as their phone number and email address. Since GPTs already have access to OpenAI's LLM, while they are integrated into ChatGPT, they do not need to explicitly make API calls to OpenAI's LLMs. Upon investigation, we found that *OpenAI Profile* was initially used as an example Action [6] in GPT creation portal [61]. *Get weather data* and *Swagger Petstore* are two such example Actions, which are embedded in 0.47% and 0.20% of the GPTs, respectively. We surmise that many developers likely unintentionally add these example Actions to their GPTs. While the inclusion of such Actions may not necessarily cause any harm to users, it shows that many GPT developers may be lay users and not experienced software developers, emphasizing the need for stricter reviewing.

We also note that several GPTs embed sophisticated Actions, such as *Zapier* [10] and *Gapier* [5], which provide dozens of APIs for various tasks, including engineering user prompts to get improved recommendations from ChatGPT. Thus, these Actions may collect an excessive amount of user data. The inclusion of such Actions may also degrade the LLM performance, as LLMs struggle with large context [44]. Other prominent Actions' functionalities include web hosting, e-commerce and shopping, and search engines.

## 4.4 Indirect Data Exposure

*4.4.1 GPTs often load several Actions.* In some cases, GPTs integrate more than one Action. Specifically, among the GPTs that integrate actions, 90.9% contain one Action, 6.6% contain two Actions, 1.2% contain three Actions, and the remaining 1.3% contain as many as 4 to 10 Actions. Of the GPTs with multiple Actions, the majority (55.3%) of them connect to additional domains (i.e., different online services), while the remaining 44.7% described other paths/endpoints for an API within the same domain (i.e., the same online service).

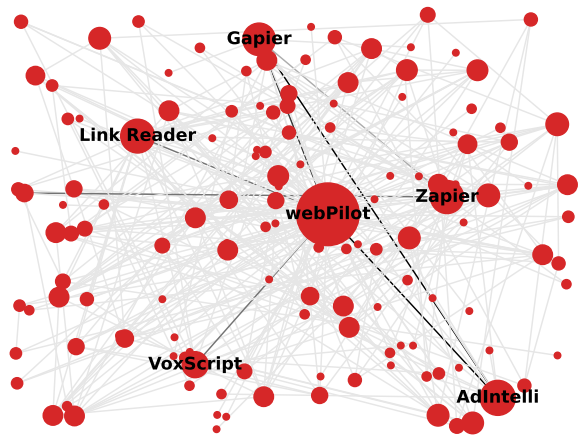

Figure 8: Action connectivity graph across GPTs. Nodes represent Actions and edges represent Action co-occurrence. Node size is proportional to its weighted degree and edge color represent its weight, i.e., edges with higher weights are darker. Nodes with a weighted degree greater than 15 are labeled with Action names.

| Policy description | % Actions |
|---|---|
| Policy of embedded services (e.g., Github, Google) | 33.5% |
| Empty policy | 27.0% |
| Actions belonging to the same vendor | 19.2% |
| JS code for dynamic rendering of privacy policy | 17.8% |
| OpenAI's privacy policy | 5.3% |
| 1x1 pixel (tracking pixel) for tracking user behavior | 3.8% |

Table 6: Description of content inside duplicate privacy policies that are seen at least 4 times.

The presence of multiple Actions can allow them to read each other's data and also influence each other's functionality, as all Actions within a GPT share the same execution space (i.e., the context window) [38, 84]. For example, as we demonstrate in our case studies represented in Figure 4 and 6, where advertising Actions co-occurring with other Actions, were able to collect additional data (including sensitive information), which was not originally intended for them.

4.4.2 *Cross-GPT inclusion of Actions.* Since Actions appear across several GPTs (Table 5), they may naturally co-occur with several other Actions, which can further increase their data exposure. We note that out of all the Actions that appear across GPTs, 23.9% co-occur with at least one other Action.

To systematically understand the potential increased exposure of data between Actions, we organize them in a co-occurrence graph. In our graph representation, nodes represent Actions and the edges represent their co-appearance in GPTs. Note that edges are undirected and weighted, such that the weight is incremented by one if the same Action pair co-occurs again in another GPT. We also make the size of a node, proportional to its weighted degree and use a color gradient to represent the edge weights, such that the darker color represents a higher weight.

Figure 8 represents the largest connected component in our graph representation. It can be seen from the figure that *webPilot* [14] and *AdIntelli* [8] Actions have the highest weighted degree in our graph (i.e., 93 and 29, respectively). Their non-weighted degrees are 63 (*webPilot*) and 12 (*AdIntelli*), which means that they co-appear with other Actions across several GPTs. In fact, we note that both *webPilot* and *AdIntelli*, co-occur in 13 GPTs. For *webPilot*, the other most frequent co-occurrences include *Gapier* [5] and *Link Reader* [7], with presence in 8 and 5 GPTs, respectively. Whereas for *AdIntelli*, the other most frequent co-occurrences include *Gapier* [5] and "Analytics to improve this assistant" [11], with presence in 9 and 3 GPTs, respectively. The presence of *AdIntelli* (an advertising service) with other "Analytics to improve this assistant" (an analytics/tracking service) seems to indicate that the LLM app ecosystem may be evolving similarly to other app ecosystems, where advertising and analytics services are often loaded together, for the purposes of targeted advertising [24, 39]. We also note that many other co-occurrences of *AdIntelli* are associated with Actions related to shopping and travel; businesses that often rely on third-party advertising and tracking services to reach their consumers. Overall, as Actions are embedded in multiple GPTs, they are in a position to connect user data collected across multiple GPTs, in different contexts.

These practices of integrating multiple Actions, especially from third parties are reminiscent of the early days of the web and mobile platforms when only a few websites/apps included a few third-party services [43]. As LLM ecosystems mature, GPTs may include dozens of Actions, including from third parties and across GPTs, as it is a common practice in the modern web, mobile, and IoT ecosystems [24, 37, 68, 80].

## 5 GPT Privacy Policy Analysis
### 5.1 Overview

OpenAI mandates individual GPT Actions to provide privacy policies, but does not require GPTs to provide a privacy policy that describes its data practices as a whole [60]. This approach deviates from the norm in other platforms, where the apps provide a privacy policy with information about their own practices, including information about services that they embed. In OpenAI's ecosystem, to understand the data practices of GPTs, users need to get familiar with the privacy policies of all of their Actions. Since the GPT interface does not disclose the Actions embedded in them, and given that Actions can dynamically embed other third-party Actions (Section 4.3.1), users may simply be unaware of the existence of these Actions in GPTs, let alone their data practices. For the purposes of analysis in this section, we analyze the privacy policy disclosures at the granularity of individual Actions.

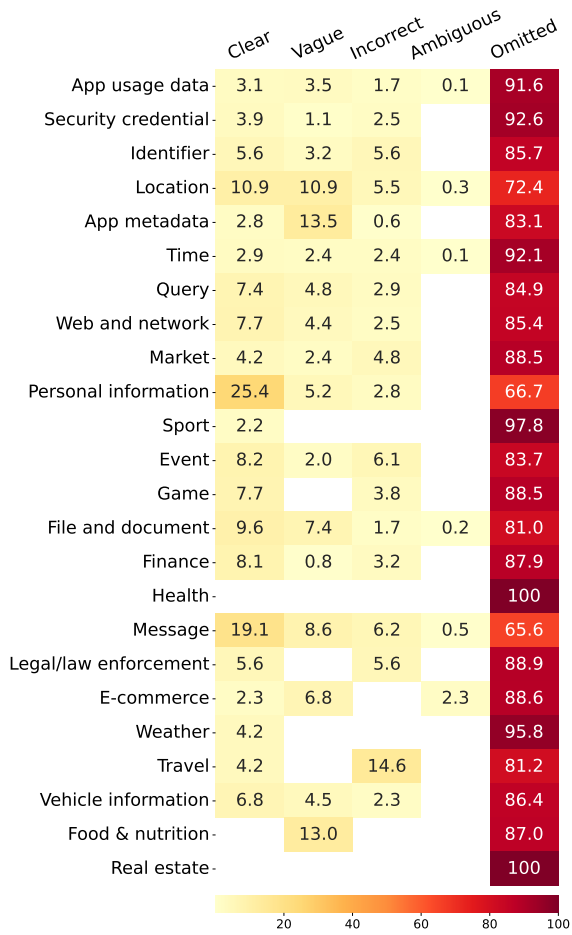

**Figure 9: Heat map of Action's data disclosure consistency by data category, for Actions which collect that data and provide privacy policies. The values represent the fraction of data for each type of disclosure. Empty cells represent the lack of respective disclosures for the respective categories.**

*5.1.1 Availability statistics.* Overall, we were able to crawl privacy policies of 93.96% of Actions (among 4,592 Actions). For the remaining Actions, the privacy policies were inaccessible. We also note that nearly 38.56% of the policies appear more than once for distinct Actions and 5.50% of the policies are near duplicates of each other (i.e., have a Jaccard similarity [1] of more than 95%).

We investigate these duplicates and near-duplicates, and provide our assessment in Table 6. We note that the inclusion of privacy policy of the external third-party services (e.g., Github, Google) is the most common reason for duplicate policies (33.5%), followed by empty privacy policies (27.0%) and Actions belonging to the same vendor (19.2%). For near-duplicates, we find that all such Actions include a boilerplate privacy policy generated from freeprivacypolicy.com, with mostly the only change being the name of the Action.

We also noted that for 12.45% of the Actions the privacy policies were less than 500 characters. We manually analyze these policies and find that they contain generic statements, such as *"We do not*

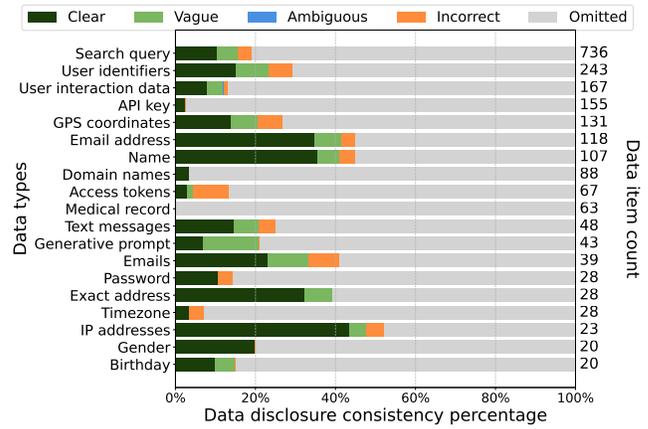

**Figure 10: Data disclosure consistency for prevalent data types (i.e., with more than 20 occurrences in Actions). Number of disclosures of each data type is shown on the right.**

*collect any personal data from users of our Service."* and *"Your data is never for sale."*. Nonetheless, they still describe the data practices of the Actions, albeit being short, thus we still consider them in our analysis.

*5.1.2 Accuracy.* Before running our framework (described in Section 3.3) at scale, we conduct a pilot study to evaluate its accuracy. We involve two human reviewers, who manually analyze the classification of data collection labels for the privacy policies of 5% (113) of unique Actions that provide privacy policies, pertaining to 625 data entities. Specifically, we check if the label assigned by our framework to a data description is correct by inspecting the relevant data collection related sentences. We consider inconsistencies (i.e., omitted, ambiguous, and incorrect) in data collection and privacy policies as positive outcomes and consistencies (i.e., clear and vague) as negative outcomes. A *true positive* occurs when an inconsistency is correctly identified, and a *false positive* when a consistent policy is incorrectly flagged as inconsistent. Similarly, a *true negative* is a correctly identified consistent data type, and a *false negative* is an inconsistent policy treated as consistent.

Overall, we achieve an accuracy of 87.44%(±1.84%) with a precision of 86.57%(±1.05%) and recall of 98.77%(±0.63%) in checking the consistency of data entities, on average across all disclosure types.

*Mistakes analysis.* After conducting a manual examination of results, we analyzed the errors associated with this task. Our analysis revealed that one-third of the errors were related to data categorized as (user) *Query*. Specifically, the privacy policies included statements such as "collect all user input and interaction data..." but our classifier failed to accurately identify this information as (user) *Query* data. Moreover, one-quarter of the errors were due to the failures in detecting paraphrased terms in the privacy policies. For example, in one of the instances, *gender* data type was referred to as *sex* in the privacy policy.

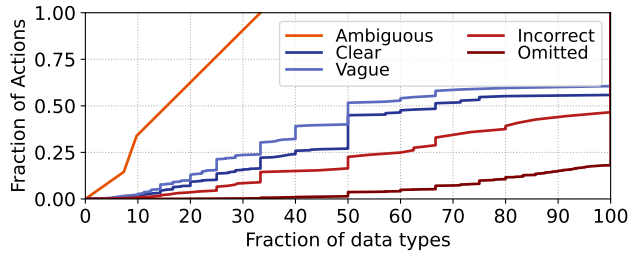

Figure 11: Distribution of data collection disclosures for Actions in their respective privacy policies.

## 5.2 Privacy Policy Analysis Results

*5.2.1 Disclosures for most data types are omitted.* Figure 9 illustrates the data disclosure consistency across all Actions and data categories and Figure 10 delves deeper into the consistency patterns for data types listed in Table 4 (we do not show data types with less than 20 occurrences). It can be seen from the figures that disclosures are omitted for most of the data categories and types. We note that for two data categories, i.e., *health* and *real estate* data, there are no disclosures. *X-Ray Analysis Service* [41] Action included in *Pneumonia Symptom Inquiry and X-Ray Analysis* GPT [42] collects users X-Ray (i.e., health-related data) but its privacy policy, is inaccessible.

Among the omitted disclosures, *Message* data category is the least omitted, followed by *personal information*, and *location*. In fact from Figure 10, we note that their data types, i.e., *Email address* and *name* belonging to *personal information* category and *exact address* belonging to *location*, are the data types that are the most clearly defined disclosures in privacy policies. For example, we note that the *Document Wizard* [21], clearly describes in its privacy policy that it: *"may collect personal information from you when you voluntarily provide it. For example, we collect your email address when you request us to send you an email with your document"* [22].

Note that the omission of disclosures is not unique to LLM apps as prior research on other platforms, such as the VR app ecosystem, found that the disclosures about the collection of most data were omitted in privacy policies [80].

*5.2.2 Nearly half of the Actions clearly disclose more than half of their data collection.* Next, we investigate whether Actions at least clearly disclose some of their data collection. Figure 11, presents the CDF of clear, vague, ambiguous, incorrect, and omitted data collection disclosures for Actions in their respective privacy policies. It can be seen from the figure that for almost half of the Actions, the data collection disclosures are consistent (i.e., clear and vague) with their privacy policies for more than half of their data collection. We also note that for nearly all Actions, at least 10% of their data collection practices are inconsistent with their disclosures.

*5.2.3 Data disclosure consistency stays almost the same as more data is collected.* We investigate, whether the consistency of disclosures decreases as Actions collect more data. Figure 12 plots the fraction of consistent data disclosures (i.e., clear and vague) over all data disclosures along with the number of collected data types by Actions. We note that as the number of collected data types increases, the consistency of disclosures only marginally decreases, however, the correlation between the two is not strong (i.e., Spearman's correlation coefficient between the two is 0.22) [73].

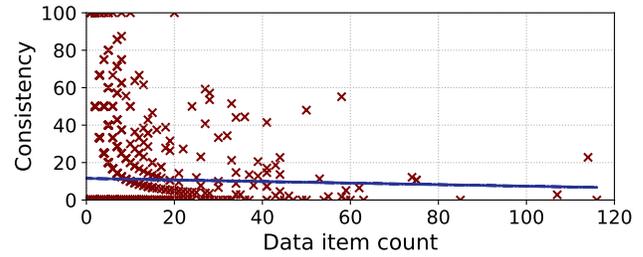

Figure 12: Fraction of consistent data disclosures (i.e., clear & vague) over all data disclosures, along with the number of collected data types by Actions. The blue line represents the underlying trend.

| Description | Clear | Vague | Total |
|---|---|---|---|
| OpenAPI definition | 0 | 20 | 20 |
| Show Me | 0 | 10 | 10 |
| Mortgage Calculator API | 8 | 0 | 8 |
| Sapientor API | 6 | 0 | 6 |
| Lowe's Product Search | 0 | 5 | 5 |
| MixerBox OnePlayer Music Plugin | 3 | 2 | 5 |

Table 7: Actions that collect more than five data types with consistent data closures in privacy policies.

We also find that the data collection of only 5.8% of Actions is consistent with their disclosures. We represent these Actions, with more five or more clear disclosures, in Table 7. Among these Action, *Mortgage Calculator* [23] and *Sapientor* [72] clearly disclose all of their data collection practices. In the case of *Sapiento*, it collects information such as the *user authentication token* and the *content provided by the user*, and clearly mentions these with the exact names in its privacy policy. In the case of *Mortgage Calculator*, it collects *loan amount* and *value of the home*, among other similar data types, and mentions in its privacy policy that it collects financial information.

## 6 Discussion

*Parallels with other emerging app ecosystems.* As compared to other ecosystems, such as the VR, Smart TVs, and Smart Speakers [37, 48, 80, 82], OpenAI's GPT app ecosystem is quickly maturing and supporting sophisticated use cases, and as a result collects a lot more data. While this data collection is enabling a wide variety of use cases, at the same time it is posing serious risks to user privacy. Considering the rapid growth of the GPT ecosystem, it is crucial that GPTs and their Actions are carefully reviewed by the vendors; which currently does not seem to be the case [17, 38, 70]. In fact, GPTs may not even be reviewed at all [58].

We also note that the LLMs provide vendors a unique opportunity to improve the privacy posture of LLM-based apps. For example, currently, OpenAI provides an interface for developers to create

GPTs using an LLM, the same LLM could also assist the GPTs in drafting their privacy policies to accurately represent their data collection practices.

*Privacy and security as key considerations in the design of LLM platforms.* LLM apps are going through a rapid transformation from providing simple instructions through a prompt, to adding dozens of third-party libraries (Actions) to support complicated use cases (Section 4.1.1). This transformation has parallels with the web ecosystem, where the websites also evolved from simple HTML web pages to complicated web applications. As a consequence, the web ecosystem suffers from chronic privacy issues, with browser vendors and researchers still continuously developing ad-hoc solutions to mitigate these concerns [29, 36, 49].

OpenAI is continuously revising its policies to catch up with the rapid growth of its app ecosystems [53, 58, 65]. However, as our measurements indicate, these efforts may not be sufficient. For example, as we note in Section 4.2.1, OpenAI requires GPTs to comply with applicable legal requirements while collecting personal user data [58, 65], but does not provide GPTs sufficient controls that they can offer to users so that users can exercise their rights. Similarly, OpenAI currently does not isolate the execution of Actions, which may lead to the indirect exposure of data between Actions embedded in a GPT (Section 4.4).

Since the LLM app ecosystem is still nascent, there is an opportunity to improve their design from the outset, instead of (and in addition to) piecemeal iterative improvements. For example, LLM app ecosystems could implement isolation-based design interfaces for multiple Actions to securely collaborate with each other inside a GPT [84]. Similarly, in addition to proposing policies, e.g., for complying with legal requirements, platforms should also develop controls so that they can be used to enforce respective policies.

*Limitations.* While we are careful in our manual assessment of our LLM-based frameworks, manual review is error-prone. To mitigate the human error biasing our results, we involve multiple humans and test a wide range of data.

We also rely on natural language descriptions of Actions' data to characterize their data practices, which may miss data collection that only occurs when Actions are executed (e.g., based on user input). However, to achieve our goal of understanding the data practices, we only need to uncover a wide range of data types, and not necessarily achieve a coverage rate of 100% (a non-trivial task even with dynamic analysis).

We also note that LLM inferences are potentially prone to errors. We conduct pilot studies that involve manual human evaluation to condition our LLM-based framework to reduce their errors. Furthermore, we quantify the error rates for our LLM-based frameworks for all of our analyses.

## 7 Conclusion

In this paper, we conducted an in-depth investigation of OpenAI's GPTs. We crawled a total of 119,274 GPTs and 4,592 Actions from third-party and the OpenAI's app stores. To characterize the practices of OpenAI's GPT Actions, we developed an in-context learning-based LLM framework. We identified that nearly half of the Actions collect 5 or more data items and one-fifth of Actions collect more than 10 data items. We also observed that GPTs embedded Actions from third-party services, which on average collected 6.03% more data than first-party Actions. We also identified several Actions that violated OpenAI's policies by collecting sensitive information, such as passwords, which are explicitly prohibited by OpenAI. Lastly, we developed an LLM-based privacy policy analysis framework to check the compliance of Action's data collection with privacy policy disclosures. We found that for most cases disclosures were omitted, with only 5.8% of Actions clearly disclosing their data collection.

## Acknowledgements

The authors would like to thank the reviewers for their valuable feedback. This work was partially supported by the NSF (CNS-2154930, CNS-2229427), ARO (W911NF-24-1-0155), ONR (N00014-24-1-2663), and Washington University in St. Louis.

## References

[1] 2011. Mining of Massive Datasets. http://infolab.stanford.edu/~ullman/mmds/ch3.pdf. http://infolab.stanford.edu/~ullman/mmds/ch3.pdf
[2] 2016. Regulation (EU) 2016/679 of the European Parliament and of the Council of 27 April 2016 on the protection of natural persons with regard to the processing of personal data and on the free movement of such data, and repealing Directive 95/46/EC (General Data Protection Regulation). https://eur-lex.europa.eu/eli/reg/2016/679/oj/eng. https://eur-lex.europa.eu/eli/reg/2016/679/oj/eng
[3] 2018. California Consumer Privacy Act. https://leginfo.legislature.ca.gov/faces/codes_displayText.xhtml?division=3.&part=4.&lawCode=CIV&title=1.81.5. https://leginfo.legislature.ca.gov/faces/codes_displayText.xhtml?division=3.&part=4.&lawCode=CIV&title=1.81.5
[4] 2023. ChatGPT Plugins: Data Exfiltration via Images & Cross Plugin Request Forgery. https://embracethered.com/blog/posts/2023/chatgpt-webpilot-data-exfil-via-markdown-injection/. https://embracethered.com/blog/posts/2023/chatgpt-webpilot-data-exfil-via-markdown-injection/
[5] 2023. Create custom versions of ChatGPT with GPTs and Zapier. https://zapier.com/blog/gpt-assistant/. https://zapier.com/blog/gpt-assistant/
[6] 2023. GPTs Example Action:"OpenAI Profile" failing on Chat Completion endpoint. https://community.openai.com/t/gpts-example-action-openai-profile-failing-on-chat-completion-endpoint/495052. https://community.openai.com/t/gpts-example-action-openai-profile-failing-on-chat-completion-endpoint/495052
[7] 2023. LinkReader - ChatGPT. https://chatgpt.com/g/g-Hdq2AC858. https://chatgpt.com/g/g-Hdq2AC858
[8] 2024. Adintelli. https://adintelli.ai/. https://adintelli.ai/
[9] 2024. AI Revenue Ops. https://www.justpaid.io/. https://www.justpaid.io/
[10] 2024. Create custom versions of ChatGPT with GPTs and Zapier. https://gapier.com/. https://gapier.com/
[11] 2024. Google Gemini Custom GPT. https://gptstore.ai/gpts/CB7_BxAKsf-google-gemini-ai. https://gptstore.ai/gpts/CB7_BxAKsf-goo-gle-gemini-ai
[12] 2024. Introducing the GPT Store. https://openai.com/blog/introducing-the-gpt-store.
[13] 2024. Provide information for Google Play's Data safety section. https://support.google.com/googleplay/android-developer/answer/10787469?hl=en.
[14] 2024. WebPilot. https://www.webpilot.ai/home?lang=en-US. https://www.webpilot.ai/home?lang=en-US
[15] Benjamin Andow, Samin Yaseer Mahmud, Justin Whitaker, William Enck, Bradley Reaves, Kapil Singh, and Serge Egelman. 2020. Actions Speak Louder than Words: Entity-Sensitive Privacy Policy and Data Flow Analysis with PoliCheck. In *29th USENIX Security Symposium (USENIX Security 20)*. USENIX Association, 985–1002. https://www.usenix.org/conference/usenixsecurity20/presentation/andow
[16] Tom B Brown, Benjamin Mann, Nick Ryder, Melanie Subbiah, Jared Kaplan, Prafulla Dhariwal, Arvind Neelakantan, Pranav Shyam, Girish Sastry, Amanda Askell, Sandhini Agarwal, Ariel Herbert-Voss, Gretchen Krueger, Tom Henighan, Rewon Child, Aditya Ramesh, Daniel M Ziegler, Jeffrey Wu, Clemens Winter, Christopher Hesse, Mark Chen, Eric Sigler, Mateusz Litwin, Scott Gray, Benjamin Chess, Jack Clark, Christopher Berner, Sam McCandlish, Alec Radford, Ilya Sutskever, and Dario Amodei. 2020. Language Models are Few-Shot Learners. In *Annual Conference on Neural Information Processing Systems (NeurIPS)*. NeurIPS.
[17] Matt Burgess. 2023. ChatGPT Has a Plug-In Problem. https://www.wired.com/story/chatgpt-plugins-security-privacy-risk/.
[18] Nicholas Carlini, Florian Tramer, Eric Wallace, Matthew Jagielski, Ariel Herbert-Voss, Katherine Lee, Adam Roberts, Tom Brown, Dawn Song, Ulfar Erlingsson,

et al. 2021. Extracting training data from large language models. In *30th USENIX Security Symposium (USENIX Security 21)*. 2633–2650.
[19] Software Freedom Conservancy. 2024. Selenium.
[20] Hao Cui, Rahmadi Trimananda, Athina Markopoulou, and Scott Jordan. 2023. PoliGraph: Automated Privacy Policy Analysis using Knowledge Graphs. In *32nd USENIX Security Symposium (USENIX Security 23)*. USENIX Association, Anaheim, CA, 1037–1054. https://www.usenix.org/conference/usenixsecurity23/presentation/cui
[21] Tati Digital. 2024. Document Wizard. https://document-wizard.com/.
[22] Tati Digital. 2024. Document Wizard - Privacy Policy. https://document-wizard.com/privacy-policy.
[23] Marcus S Elola. 2024. ChatGPT - Mortgage Calculator. https://chatgpt.com/g/g-NIGpQi8Rc.
[24] Steven Englehardt and Arvind Narayanan. 2023. Online Tracking: A 1-million-site Measurement and Analysis. (2023).
[25] Bubeck et. al. 2023. Sparks of Artificial General Intelligence: Early experiments with GPT-4. https://arxiv.org/abs/2303.12712. (2023).
[26] OpenAI Forum. 2023. Is there a definitive list of all GPTs on the store?
[27] Tianyu Gao, Adam Fisch, and Danqi Chen. 2021. Making Pre-trained Language Models Better Few-shot Learners. In *Proceedings of the 59th Annual Meeting of the Association for Computational Linguistics and the 11th International Joint Conference on Natural Language Processing (Volume 1: Long Papers)*. 3816–3830.
[28] Amir Globerson, Gal Chechik, Fernando Pereira, and Naftali Tishby. 2004. Euclidean embedding of co-occurrence data. *Advances in neural information processing systems* 17 (2004).
[29] Google. [n. d.]. Google Privacy Sandbox. https://privacysandbox.com.
[30] Google. 2018. Improve extension security. https://developer.chrome.com/docs/extensions/develop/migrate/improve-security.
[31] Google. 2023. Google Gemini. https://gemini.google.com/.
[32] Google. 2023. Trustworthy chrome extensions. https://blog.chromium.org/2018/10/trustworthy-chrome-extensions-by-default.html.
[33] Kai Greshake, Sahar Abdelnabi, Shailesh Mishra, Christoph Endres, Thorsten Holz, and Mario Fritz. 2023. Not what you've signed up for: Compromising Real-World LLM-Integrated Applications with Indirect Prompt Injection. *arXiv preprint arXiv:2302.12173* (2023).
[34] IETF OAuth Working Group. 2023. The OAuth 2.1 Authorization Framework. https://www.ietf.org/archive/id/draft-ietf-oauth-v2-1-09.html.
[35] Hamza Harkous, Kassem Fawaz, Rémi Lebret, Florian Schaub, Kang G. Shin, and Karl Aberer. 2018. Polisis: Automated Analysis and Presentation of Privacy Policies Using Deep Learning. arXiv:1802.02561 [cs.CL]
[36] Apple Inc. 2024. WebKit Tracking Prevention Policy. https://webkit.org/tracking-prevention-policy/.
[37] Umar Iqbal, Pouneh Nikkhah Bahrami, Rahmadi Trimananda, Hao Cui, Alexander Gamero-Garrido, Daniel Dubois, David Choffnes, Athina Markopoulou, Franziska Roesner, and Zubair Shafiq. 2023. Tracking, Profiling, and Ad Targeting in the Alexa Echo Smart Speaker Ecosystem. In *ACM Internet Measurement Conference (IMC)*.
[38] Umar Iqbal, Tadayoshi Kohno, and Franziska Roesner. 2024. LLM Platform Security: Applying a Systematic Evaluation Framework to OpenAI's ChatGPT Plugins. In *Proceedings of the 2024 AAAI/ACM Conference on AI, Ethics, and Society (AIES)*.
[39] Umar Iqbal, Charlie Wolfe, Charles Nguyen, Steven Englehardt, and Zubair Shafiq. 2022. Khaleesi: Breaker of advertising and tracking request chains. In *31st USENIX Security Symposium (USENIX Security 22)*. 2911–2928.
[40] Muhammad Junaid. 2024. Cal AI. https://caxgpt.vercel.app.
[41] Harsha Vardhan Khurdula. 2024. KhurdhulaHarshavardhan X-ray Analysis Service. https://khurdhulaharshavardhan-jhvvqrbzyq-uc.a.run.app.
[42] Harsha Vardhan Khurdula. 2024. Pneumonia Symptom Inquiry and X-Ray Analysis GPT. https://chatgpt.com/g/g-OMMTGHdSv-pneumonia-symptom-inquiry-and-x-ray-analysis-gpt.
[43] Ada Lerner, Anna Kornfeld Simpson, Tadayoshi Kohno, and Franziska Roesner. 2016. Internet jones and the raiders of the lost trackers: An archaeological study of web tracking from 1996 to 2016. In *25th USENIX Security Symposium (USENIX Security 16)*.
[44] Tianle Li, Ge Zhang, Quy Duc Do, Xiang Yue, and Wenhu Chen. 2024. Long-context LLMs Struggle with Long In-context Learning. *arXiv preprint arXiv:2404.02060* (2024).
[45] Zengrui Liu, Umar Iqbal, and Nitesh Saxena. 2024. Opted Out, Yet Tracked: Are Regulations Enough to Protect Your Privacy?. In *Privacy Enhancing Technologies Symposium (PETS)*.
[46] Nils Lukas, Ahmed Salem, Robert Sim, Shruti Tople, Lukas Wutschitz, and Santiago Zanella-Béguelin. 2023. Analyzing leakage of personally identifiable information in language models. In *2023 IEEE Symposium on Security and Privacy (SP)*. IEEE, 346–363.
[47] Célestin Matte, Nataliia Bielova, and Cristiana Santos. 2020. Do cookie banners respect my choice?: Measuring legal compliance of banners from iab europe's transparency and consent framework. In *2020 IEEE Symposium on Security and Privacy (SP)*. IEEE, 791–809.
[48] Hooman Mohajeri Moghaddam, Gunes Acar, Ben Burgess, Arunesh Mathur, Danny Yuxing Huang, Nick Feamster, Edward W. Felten, Prateek Mittal, and Arvind Narayanan. 2019. Watching You Watch: The Tracking Ecosystem of Over-the-Top TV Streaming Devices. In *Proceedings of the 2019 ACM SIGSAC Conference on Computer and Communications Security* (London, United Kingdom) *(CCS '19)*. Association for Computing Machinery, New York, NY, USA, 131–147.
[49] Shaoor Munir, Patrick Lee, Umar Iqbal, Zubair Shafiq, and Sandra Siby. 2024. PURL: Safe and Effective Sanitization of Link Decoration. In *USENIX Security Symposium*.
[50] NLTK. [n. d.]. NLTK Tokenization. https://www.nltk.org/api/nltk.tokenize.html.
[51] Harsha Nori, Yin Tat Lee, Sheng Zhang, Dean Carignan, Richard Edgar, Nicolo Fusi, Nicholas King, Jonathan Larson, Yuanzhi Li, Weishung Liu, et al. 2023. Can generalist foundation models outcompete special-purpose tuning? case study in medicine. *arXiv preprint arXiv:2311.16452* (2023).
[52] OpenAI. 2022. Introducing ChatGPT. https://openai.com/blog/chatgpt.
[53] OpenAI. 2023. Actions in production. https://platform.openai.com/docs/actions/production.
[54] OpenAI. 2023. Can I charge people money for my plugin? https://community.openai.com/t/exploring-ways-to-monetize-free-chatgpt-plugins/331899.
[55] OpenAI. 2023. Creating a GPT - OpenAI. https://help.openai.com/en/articles/8554397-creating-a-gpt
[56] OpenAI. 2023. Data Controls FAQ. https://help.openai.com/en/articles/7730893-data-controls-faq.
[57] OpenAI. 2023. Getting started. https://help.openai.com/en/articles/8554402-gpts-data-privacy-faqs.
[58] OpenAI. 2023. Plugins and Actions Terms. https://openai.com/policies/plugin-terms/.
[59] OpenAI. 2024. Getting Started with Actions. https://platform.openai.com/docs/actions/getting-started. Accessed: 2024-06-07.
[60] OpenAI. 2024. GPT Actions. https://platform.openai.com/docs/actions/introduction.
[61] OpenAI. 2024. GPT Editor. https://chatgpt.com/gpts/editor.
[62] OpenAI. 2024. GPTs. https://chatgpt.com/gpts.
[63] OpenAI. 2024. Introducing GPTs. https://openai.com/blog/introducing-gpts.
[64] OpenAI. 2024. Memory and new controls for ChatGPT. https://openai.com/index/memory-and-new-controls-for-chatgpt/
[65] OpenAI. 2024. Usage Policies. https://openai.com/policies/usage-policies.
[66] OpenAI. 2024. Usage Policies. https://openai.com/policies/usage-policies/.
[67] Panagiotis Papadopoulos, Nicolas Kourtellis, and Evangelos P. Markatos. 2019. Cookie Synchronization: Everything You Always Wanted to Know But Were Afraid to Ask. In *The Web Conference (WWW)*.
[68] Abbas Razaghpanah, Rishab Nithyanand, Narseo Vallina-Rodriguez, Srikanth Sundaresan, Mark Allman, Christian Kreibich, Phillipa Gill, et al. 2018. Apps, trackers, privacy, and regulators: A global study of the mobile tracking ecosystem. In *The 25th Annual Network and Distributed System Security Symposium (NDSS 2018)*.
[69] Reddit. 2024. There are already 51 unofficial GPT stores being discovered.
[70] Johann Rehberger. 2023. Plugin Vulnerabilities: Visit a Website and Have Your Source Code Stolen. https://embracethered.com/blog/posts/2023/chatgpt-plugin-vulns-chat-with-code/. https://embracethered.com/blog/posts/2023/chatgpt-plugin-vulns-chat-with-code/
[71] N Reimers. 2019. Sentence-BERT: Sentence Embeddings using Siamese BERT-Networks. *arXiv preprint arXiv:1908.10084* (2019).
[72] Sapientor.net. 2024. ChatGPT - Knowledge Base GPT. https://chatgpt.com/g/g-rGJvqSptw.
[73] Patrick Schober, Christa Boer, and Lothar A Schwarte. 2018. Correlation coefficients: appropriate use and interpretation. *Anesthesia & analgesia* 126, 5 (2018), 1763–1768.
[74] Muhammad Junaid Shaukat. 2024. Cax TaskPal. https://chatgpt.com/g/g-fC8sZoDCi-cax-taskpal.
[75] Robin Staab, Mark Vero, Mislav Balunovic, and Martin Vechev. 2024. Beyond Memorization: Violating Privacy via Inference with Large Language Models. In *The Twelfth International Conference on Learning Representations*. https://openreview.net/pdf?id=kmn0BhQk7p/
[76] Dongxun Su, Yanjie Zhao, Xinyi Hou, Shenao Wang, and Haoyu Wang. 2024. Gpt store mining and analysis. *arXiv preprint arXiv:2405.10210* (2024).
[77] Swagger Group. 2024. OpenAPI Spec 3.1.0.
[78] Zhaoxuan Tan and Meng Jiang. 2023. User Modeling in the Era of Large Language Models: Current Research and Future Directions. arXiv:2312.11518 [cs.CL]
[79] TechCrunch. 2023. Google launches a smarter Bard. https://techcrunch.com/2023/05/10/google-launches-a-smarter-bard/.
[80] Rahmadi Trimananda, Hieu Le, Hao Cui, Janice Tran Ho, Anastasia Shuba, and Athina Markopoulou. 2022. {OVRseen}: Auditing network traffic and privacy policies in oculus {VR}. In *31st USENIX security symposium (USENIX security 22)*. 3789–3806.
[81] u/AwkwardAsHell. 2023. This is scary! Posting stuff by itself. - Reddit. https://www.reddit.com/r/OpenAI/comments/146xl6u/comment/jqt6ezb/.

[82] Janus Varmarken, Hieu Le, Anastasia Shuba, Athina Markopoulou, and Zubair Shafiq. 2020. The tv is smart and full of trackers: Measuring smart tv advertising and tracking. *Proceedings on Privacy Enhancing Technologies* (2020).
[83] Jason Wei, Xuezhi Wang, Dale Schuurmans, Maarten Bosma, Fei Xia, Ed Chi, Quoc V Le, Denny Zhou, et al. 2022. Chain-of-thought prompting elicits reasoning in large language models. *Advances in neural information processing systems* 35 (2022), 24824–24837.
[84] Yuhao Wu, Franziska Roesner, Tadayoshi Kohno, Ning Zhang, and Umar Iqbal. 2025. IsolateGPT: An execution isolation architecture for llm-based systems. In *Network and Distributed System Security Symposium (NDSS)*.
[85] Chuan Yan, Ruomai Ren, Mark Huasong Meng, Liuhuo Wan, Tian Yang Ooi, and Guangdong Bai. 2024. Exploring ChatGPT App Ecosystem: Distribution, Deployment and Security. In *IEEE/ACM International Conference on Automated Software Engineering*.
[86] Zejun Zhang, Li Zhang, Xin Yuan, Anlan Zhang, Mengwei Xu, and Feng Qian. 2024. A first look at gpt apps: Landscape and vulnerability. *arXiv preprint arXiv:2402.15105* (2024).
[87] Wanru Zhao, Vidit Khazanchi, Haodi Xing, Xuanli He, Qiongkai Xu, and Nicholas Donald Lane. 2024. Attacks on Third-Party APIs of Large Language Models. In *ICLR Workshop on Secure and Trustworthy Large Language Models*.
[88] Yanjie Zhao, Xinyi Hou, Shenao Wang, and Haoyu Wang. 2024. Llm app store analysis: A vision and roadmap. *arXiv preprint arXiv:2404.12737* (2024).

## A  Ethics

We identify several privacy issues and problematic GPTs and Actions in the OpenAI's ecosystem, which could pose harm to the users. We disclose all such instances to OpenAI and relevant third-party developers, so that they can address the uncovered issues.

As OpenAI does not provide an API to crawl their GPT store, we did not try to reverse engineer it, and instead used publicly listed GPTs on third-party stores to conduct our analysis.

## B  GPT Manifest and Action Specification

GPTs in the OpenAI ecosystem are similar to apps in the Android or the iOS mobile platforms. GPTs consist of two main components: (i) a manifest that describes the functionality provided by the GPT in natural language, and (ii) a set of built-in and custom tools (referred to as Actions). Actions are implemented as web APIs, and exposed to GPTs via specification files that list the API endpoints and their textual descriptions. Actions can be considered modular extensions that allow GPTs to interact with external online services via APIs, thereby enabling GPTs to retrieve real-time data or execute operations beyond the LLM's intrinsic capabilities.

While GPTs serve as the primary interface for end-users, Actions function as programmable endpoints that augment GPT capabilities through developer-defined logic and authenticated access. When a user enables and uses a GPT, GPT manifest, built-in tools, and Action specifications are loaded in the context window of a dedicated LLM instance. Based on the user query, GPTs dynamically invoke Actions if needed.

We next explain GPT manifests and Action specifications in detail.

### B.1  GPT Manifest

The JSON manifests of GPTs describe their functionality in natural language, including the endpoints contacted by Actions, and the data collected by them. Code 1 describes a simplified manifest of a custom GPT, named "Ultimate Travel Planner", that aims to plan travel for users. In the manifest, the *display* field contains information about the GPT submitted by the author; this includes a name, description, and suggested prompts for interacting with the GPT. The `tools` field contains an array of JSON objects, where each object is a tool with a field called `type` that indicates what kind of tool is enabled (e.g., DALL-E, browser). The exception to this rule is Actions, which also contain a *metadata* field which includes important information about the Action, such as its privacy policy, domain, and Action specification in OpenAPI format (described next). Moreover, there is a `files` field in the GPT manifest which indicates files used by the GPT as its knowledge base. One file is uploaded in this example, but we are only able to see the MIME type and an ID that is specific to the GPT.

Besides the *display*, *tools*, and *files* fields, a GPT manifest also contains an `id` field, which is a unique 10-character alphanumeric shortcode that identifies the GPT and is used as the short link to access the GPT. Also included is a *tags* field, which tags GPTs with important attributes about the GPT. We observe that OpenAI has used these tags to identify GPTs: `first_party`, `public`, `private`, `reportable`, `unreviewable`, and `uses_function_calls`).

### B.2  Action Specification

Code 2 shows an expanded view of the OpenAPI specification used in the "Ultimate Travel Planner" GPT. This action uses a third-party KAYAK API to search for deals on flights, hotels, and cars. The composition of an OpenAPI specification can differ, but as a standard rule, OpenAPI specifications contain at least a `servers`, `info`, `paths`, and `OpenAPI` field which respectively denote the URLs hosting the API, an overview of the specification, the endpoint locations, and version of the OpenAPI specification used [77].

## C  LLM Prompts

### C.1  Prompts for Data Collection Analysis

The prompts used in the proposed LLM-based framework for data description classification and addressing non-classified data descriptions can be found in Code 3 and Code 4, respectively.

### C.2  Prompts for Privacy Policy Analysis

The prompts used in the proposed LLM-based framework for identifying data collection-related sentences and assigning data collection consistency labels can be found in Code 5 and Code 6, respectively.

## D  GPT Data Taxonomy

Table 8 lists all categories and data types included in the final constructed data taxonomy.

```
 1  {
 2      "gizmo": {
 3          "id": "g-fYBGstD4a",
 4          "author": {
 5              "display_name": "Stephan B\u00fcttig",
 6          },
 7          "display": {
 8              "name": "\u2708\ufe0f Ultimate Travel Planner (5.0\u2b50)",
 9              "description": "Plan your perfect trip with this Ultimate Travel Planner! This multilingual and versatile
                   GPT will help you create a customized itinerary based on your chosen destination, duration, type of
                   vacation, and means of travel.",
10              "prompt_starters": [
11                  "Plan a surprise trip for me to any specific destination.",
12                  ...
13              ]
14              "categories": ["productivity"]
15          },
16          "tags": ["public", "reportable", "uses_function_calls"],
17      },
18      "tools": [
19          {
20              "id": "kPT04JXwhSVSVhIAmfMhWCcd",
21              "type": "action(plugins_prototype)"
22              "json_spec": { see listing 2 }
23          },
24          {
25              "type": "browser",
26          }
27          {
28              "type": "dalle",
29          }
30      ],
31      "files": [
32          {
33              "id": "gzm_cnf_iyezMwP5wAgnulFCIngNGhSI~gzm_file_syV7QUY1KycmdLnN3v4FcaMb",
34              "type": "",
35          }
36      ]
37  }
```

**Code 1: A simplified representation of the *Ultimate Travel Planner*, a custom GPT designed to assist users in planning trips. It leverages various capabilities of OpenAI's platform, including uploaded files, web browsing, Actions, and DALL-E.**

```json
{
    "openapi": "3.0.1",
    "info": {
        "title": "KAYAK - Flights, Hotels, Cars",
        "description": "A plugin that allows users to search for the best deals on flights, hotels and cars",
        "version": "v1"
    },
    "servers": [{"url": "https://www.kayak.com",}],
    "paths": {
        "/sherlock/aiplugin/search/flights": {
            "post": {
                "summary": "Search flights on a flight route for certain dates",
                "x-openai-isConsequential": false,
                "requestBody": {
                    "content": {
                        "application/json": {
                            "schema": ..
                        }
                    }
                }
                "parameters" : [
                    {
                        "name": "format",
                        "in": "query",
                        "required": true,
                        "schema": {
                            "type": "string",
                            "default": "json"
                        }
                        "description": "The format of the response."
                    }
                ],
                "responses": {
                    "200": {"description": "OK"},
                    "429": {"description": "Rate limited"}
                }
            }
        },
        ...
    }
}
```

Code 2: An expanded OpenAPI specification for Ultimate Travel Planner's Action which specifies a third-party API that allows users to search for the best deals on flights, hotels, and cars.

```
Objective:
You are a data classification assistant. Your objective is to categorize each data entity into ONE data type within this data taxonomy. For
data entities not covered by the taxonomy, you should categorize them as "Other".

The data taxonomy is as follows:
{initial_taxonomy}

There are some additional criteria for the classification task:
{categorization_rules}

Here are some additional examples for your reference:
{general_examples}

Here is the format of data entity provided for each query:
[
    "name and description": "Name and description of the data entity",
    "examples": [...],
]

For understanding each data entity:
1. Read the name and description of the data entity. Note: ***You should consider the entire description, not just a part of it.***
2. Review the examples attached to the data entity to determine if any are similar to the data sample.
3. If you still believe the data entity does not belong to any data type, classify it as "Other."

You should follow the steps below to categorize each data entity:
1. You need to fully understand the data taxonomy and follow the additional criteria mentioned below. Be sure to refer to the description of
each data type for better understanding. You are not allowed to identify the data types based solely on their names.
2. Read all the information provided in the input.
3. Review all the examples attached, and ask yourself, "Do any of the examples have the same meaning as this data entity?" If so, the label of
the example is highly likely to be the correct label for the data entity.
4. Categorize the current data entity into one data type.
5. Double-check your answer by asking yourself, "Is this data entity covered by the description of the chosen data type?"
6. If you are confident in your answers, you can submit them. Otherwise, go back to the previous step, revise your answers, or consider the
"Other" label.

Note: you MUST select the most appropriate data type for each data entity.
There are multiple data entities in the input, and you need to categorize each of them independently.
Follow the output example below:
{output_format}

You MUST STRICTLY follow the provided output example. Respond only in the specified JSON format, with no additional text.
```

**Code 3: Prompt for data description classification.**

```
Objective:
You are a data taxonomy expert. Your objective is to decide whether the data entities are valuable to create a new sub datatype for it and add
it to the existing data taxonomy.
Attention: We want a concise data taxonomy instead of a comprehensive one. This will affect your decision!
The existing data taxonomy is as follows:
{existing_taxonomy}

Here is the format of data entity:
[
    "name : description": "data_entity_name : data_entity_description",
    "amount appears": int,
]

Note: For whether a data entity is valuable, you should consider the following aspects:
1. It should be distinct from the existing sub data types.
2. It should provide additional value to the existing data taxonomy.
3. Bigger amount appears means the data entity is more valuable, as it appears more frequently in the dataset.

For each data entities, choose one action from the following options:
1. ['Covered', 'An existing sub datatype']. If the data entity is covered by an existing sub data type, choose this option and specify the
existing sub data type.
2. ['Add', 'New sub datatype']. If the data entity is valuable and should create a new sub datatype for it, choose this option.
3. ['Combine', 'New sub datatype']. If the data entity is valuable but should be combined with other data entity to create a new sub datatype,
choose this option and specify new sub data type. You can rewrite the description if necessary.
4. ['Deprecate', '']. If the data entity is not valuable and should be deprecated, choose this option.

After you have a draft of the new data taxonomy and decisions for each data entity, you need to check whether you can improve it by [Deprecate]
less valuable new sub datatypes or [Combine] new sub data types to make the data taxonomy more concise. Make sure to update all the decisions
accordingly.

Follow the output example below, make sure output all the decisions for each data entity in the same order as the input:
{output_format}

You MUST STRICTLY follow the above provided output example. Only answer with the specified JSON format, no other text.
```

**Code 4: Prompt for addressing non-classified data descriptions.**

```
Objective:
You are a privacy policy data collection statement extractor.

You will be given sentences from privacy policy and your goal is to identify sentences related to data collection.

Some example data collection:

{collection_statement_examples}

Create an output represented in JSON containing the following:
{output_format}

You MUST STRICTLY follow the above provided output example. Only answer with the specified JSON format, no other text.
```

**Code 5: Prompt for identifying data collection-related sentences.**

```
Objective:
You are a privacy policy consistency checker. You analyze whether apps disclose their data collection practices in their privacy policies.

You will be given a list of sentences related to data collection from an app's privacy policy as well as the information of a data entity
disclosed by the same app. Your goal is to examine whether the data collection and sharing practices of the data entity is mentioned in the
privacy policy by assigning one of the following labels for each sentence:

    CLEAR: If the data type description exactly matches a data type in the collection statement
    VAGUE: If the data type description is mentioned in broader or vague terms in the collection statement
    AMBIGUOUS: If there are contradictory collection statements about a data type description. For example if one statement says that the data
is collected and another statement states that the data is not collected.
    INCORRECT: If the data types is collected and the statements say that the data is not collected
    OMITTED: If the data collection statements do not mention the collected data type at all

Here are some examples to help you assign the appropriate labels:

{examples}

Create an output represented in JSON containing the following for each data type description.
Note that you MUST select the most appropriate label for each sentence and data entity pair and follow the output example below:
{output_format}

You MUST STRICTLY follow the above provided output example and generate a label for each sentence. Only answer with the specified JSON format,
no other text.
```

**Code 6: Prompt for assigning data collection consistency labels.**

| Category | Data types |
| --- | --- |
| Location | Altitude, Exact address, City, Street, State/province, Country, Postcode, Place of interest, GPS coordinates, Relative location, Route, General location, Origin/destination, Region |
| Time | Year, Time period, Season, Month, Week, Time of day, Date, Relative time, Timezone, Frequency, Timestamp |
| Event information | Event name, Event description, Participants, Reminders |
| Personal information | Relationship, Age, Birthday, Race and ethnicity, Sexual orientation, Name, Gender, Education, Work, Email address, Phone number, Social media handle, Mailing address, Nickname |
| Finance information | Purchase history, Insurance, Property ownership, Loans, Income information, Investment |
| Health information | Medical record, Fitness information |
| App usage data | Status, Subscription information, Diagnostics, Current session setting, Response fields, User interaction data |
| App metadata | Function description, Name or version, Publisher, Integrated applications |
| Files and documents | File path, File name, File hash, File type, File description, File size, File content, Source, File list |
| Web and network data | URLs, IP addresses, Domain names, Related links, Connection logs, Blockchain data, Cookies, Web page content, User-Agent strings, Database information, Multimedia data |
| Message | Text messages, Emails, Participants, User feedback |
| Query | Query filter, Generative prompt, Search query |
| Identifier | Vehicle identification number (VIN), License plate number, Device IDs, Resource IDs, Project and issue identifiers, Account identifiers, Media identifiers, Geographical area codes, Financial instrument identifiers, Product and item identifiers, Ticket and order identifiers, Organization identifiers, User identifiers |
| Market data | Ticker symbol, Company name, Exchange, List of ticker symbols, Currency information, Financial ratios and metrics |
| Weather information | Weather data parameters, Weather data timeframe |
| Vehicle information | Vehicle make, Vehicle model, Vehicle type, Vehicle color, Vehicle mileage, Vehicle fuel type, Vehicle specifications |
| Security credentials | API key, Password, Access tokens, Cryptographic key, Verification code |
| Food and nutrition information | Nutrients, Recipes, Food type filters, Meal planning |
| Real estate data | Property details, Amenities, Furnishing status |
| E-commerce data | Parcel dimensions, Product details, Company information, Business metrics, E-commerce transaction details |
| Gaming data | In-game data, Player statistics |
| Legal and law enforcement data | Crime details, Case outcomes and evidence, Legal provisions, Legal inquiries |
| Travel information | Baggage information, Cabin preferences, Passenger counts |
| Sports information | Markets, Teams, Leagues, Statistics |

Table 8: Overview of categories and data types in the final data taxonomy.